# Multi-party Quantum Private Comparison Based on the Entanglement Swapping of $d$-level Cat States and $d$-level Bell states

Zhao-Xu Ji, Tian-Yu Ye *

College of Information & Electronic Engineering, Zhejiang Gongshang University, Hangzhou 310018, P.R.China
*E-mail：happyyty@aliyun.com

**Abstract:** In this paper, a novel multi-party quantum private comparison (MQPC) protocol with a semi-honest third party (TP) is proposed based on the entanglement swapping of $d$-level cat states and $d$-level Bell states. Here, TP is allowed to misbehave on his own but will not conspire with any party. In our protocol, $n$ parties employ unitary operations to encode their private secrets and can compare the equality of their private secrets within one time execution of the protocol. Our protocol can withstand both the outside attacks and the participant attacks on the condition that none of the QKD methods is adopted to generate keys for security. One party cannot obtain other parties' secrets except for the case that their secrets are identical. The semi-honest TP cannot learn any information about these parties' secrets except the end comparison result on whether all private secrets from $n$ parties are equal.

## 1 Introduction

Since the pioneering work of quantum cryptography was proposed by Bennett and Brassard [1] in 1984, various secure quantum cryptography protocols have been proposed, such as quantum key distribution (QKD) [2-5], quantum secure direct communication (QSDC) [6-8], quantum teleportation [9-11] and so on.

Secure multi-party computation (SMC), which is a fundamental primitive in modern cryptography, is to compute a function with private inputs from different parties in a distributed network without disclosing the genuine content of each private input. It has wide applications in private bidding and auctions, secret ballot elections, e-commerce and data mining and so on, and has long been the object of intensive study in classical cryptography. However, the security of SMC is based on the computation complexity assumption, which is susceptible to the strong ability of quantum computation. Under this circumstance, the classical SMC has been generalized into its counterpart in the realm of quantum mechanics, i.e., quantum secure multi-party computation (QSMC).

Private comparison, first introduced by Yao [12] in the millionaires' problem, is a basic problem of SMC. In the millionaires' problem [12], two millionaires wish to judge who is richer without knowing the actual property of each other. Afterward, Boudot et al. [13] proposed a protocol to decide whether two millionaires are equally rich or not. However, Lo [14] pointed out that it is impossible to construct a secure equality function in a two-party scenario. Hence, some additional assumptions, such as a third party (TP), should be taken into account for accomplishing the goal of private comparison.

Quantum private comparison (QPC), which can be seen as the generalization of classical private comparison into the realm of quantum mechanics, was first proposed by Yang and Wen [15] in 2009. The goal of QPC is to judge whether the private inputs from different parties are equal or not by utilizing the principles of quantum mechanics on the basis that none of their genuine contents are leaked out. Since then, many QPC protocols [16-38] have been constructed.

Multi-party quantum private comparison (MQPC) is the QPC where there are more than two parties who want to compare the equality of their private inputs. Supposed that there are $n$ parties, if they adopt a two-party QPC protocol to accomplish the equality comparison of their private inputs, they will have to implement this protocol with $(n-1) \sim n(n-1)/2$ times. As a result, the comparison efficiency is reduced. A question naturally arouses: can the equality comparison of private inputs from $n$ parties be achieved within one execution of protocol? In 2013, Chang et al. [39] suggested the first MQPC protocol by using $n$-particle GHZ class states to give this question a positive answer, which can accomplish arbitrary pair's comparison of equality among $n$ parties within one execution. Since then, MQPC has attracted more and more attentions. The MQPC protocols with multi-level quantum system [40,41] were also constructed afterward.

Based on the above analysis, in this paper, we propose a novel MQPC protocol, which utilizes the entanglement swapping of $d$-level cat states and $d$-level Bell states. Our protocol can accomplish the equality comparison of secrets from $n$ parties within one time execution. One party cannot obtain other parties' secrets except for the case that their secrets are identical. The semi-honest TP cannot learn any information about these parties' secrets except the end comparison result on whether all private secrets from $n$ parties are equal.

The rest of this paper is organized as follows. In Sect.2, we introduce the preliminary knowledge used in our protocol. In Sect.3, we describe our protocol for three parties and for qubits first, and then generalize it to the version for $n$ parties and for qudits. In Sect.4, we analyze our protocol from the aspects on output correctness and security. Finally, discussion and conclusion are given in Sect.5.

## 2 Preliminary knowledge

Before describing our protocol, it is necessary to introduce the preliminary knowledge used in our protocol.

In $d$-level system (qudits), for the $d$-level basis states $|k\rangle$ ($k \in \{0,1,\ldots,d-1\}$), the $d$ th order discrete fourier transform is defined as

$$F|k\rangle = \frac{1}{\sqrt{d}}\sum_{r=0}^{d-1}\zeta^{kr}|r\rangle, \qquad (1)$$

where $\zeta = e^{2\pi i/d}$. The two sets, $V_1 = \{|k\rangle\}_{k=0}^{d-1}$ and $V_2 = \{F|k\rangle\}_{k=0}^{d-1}$, construct two common non-orthogonal bases.

A generalization of the familiar Bell states for qudits introduced in Refs.[42-44] is a set of $d^2$ maximally entangled states which form an orthonormal basis for the space of two qudits. Their explicit forms are

$$|\Psi(u,v)\rangle = \frac{1}{\sqrt{d}}\sum_{j=0}^{d-1}\zeta^{ju}|j\rangle|j+v\rangle, \qquad (2)$$

where labels $u$ and $v$ run from 0 to $d-1$. Easily, one can get

$$|\Psi(0,0)\rangle = \frac{1}{\sqrt{d}}\sum_{j=0}^{d-1}|j\rangle|j\rangle. \qquad (3)$$

One can generate any Bell state $|\Psi(u,v)\rangle$ by acting on $|\Psi(0,0)\rangle$ with $|U_{(u,v)}\rangle$ which is expressed as

$$|U_{(u,v)}\rangle = \sum_{j=0}^{d-1}\zeta^{ju}|j+v\rangle\langle j|, \qquad (4)$$

i.e.,

$$\left(I \otimes |U_{(u,v)}\rangle\right)|\Psi(0,0)\rangle = |\Psi(u,v)\rangle. \qquad (5)$$

The $d$-level $n$-particle cat states introduced in Ref.[45] have the form

$$|\Psi(u_1,u_2,\ldots,u_n)\rangle = \frac{1}{\sqrt{d}}\sum_{j=0}^{d-1}\zeta^{ju_1}|j,j+u_2,j+u_3,\ldots,j+u_n\rangle, \qquad (6)$$

where labels $u_1,u_2,\ldots,u_n$ run from 0 to $d-1$. These cat states are orthonormal and complete.

As mentioned in Ref.[45], the formula of entanglement swapping between one $d$-level Bell state $|\Psi(v,v')\rangle_{s,s'}$ and one $d$-level $n$-particle cat state $|\Psi(u_1,u_2,\ldots,u_n)\rangle_{1,2,\ldots,n}$, where the entanglement swapping does not involve the first particle of cat state (i.e., $m \in \{2,3,\ldots,n\}$), can be expressed as

$$|\Psi(u_1,u_2,\ldots,u_n)\rangle_{1,2,\ldots,n} \otimes |\Psi(v,v')\rangle_{s,s'} = \frac{1}{d}\sum_{k,l=0}^{d-1}\zeta^{kl}|\Psi(u_1+k,u_2,u_3,\ldots,v'+l,\ldots,u_n)\rangle_{1,2,\ldots,s',\ldots,n} \otimes |\Psi(v-k,u_m-l)\rangle_{s,m}. \qquad (7)$$

The graphical description of formula (7) is given in Fig.1 [45]. Here, one $d$-level $n$-particle cat state is depicted by a line with $n$ nodes on it where the first node is solid and other nodes are empty. And one $d$-level Bell state is depicted by a line with two nodes on it where one node is solid and the other node is empty. The simple rule on labels of these two states during this entanglement swapping is

$$\begin{cases} v \to v-k \\ u_1 \to u_1+k \end{cases}, \quad \begin{cases} v' \to v'+l \\ u_m \to u_m-l \end{cases}, \quad u_i \to u_i \text{ for } i=2,3,\ldots,m-1,m+1,\ldots,n. \qquad (8)$$

## 3 The proposed MQPC protocol
### 3.1 The proposed MQPC protocol for three parties and for qubits

Suppose that there are three parties named $P_1, P_2, P_3$, where $P_i$ ($i=1,2,3$) has a private secret $x_i$. The binary representation of $x_i$ in $F_{2^L}$ is $(x_i^1, x_i^2, \ldots, x_i^L)$, where $x_i^j \in \{0,1\}$ for $j=1,2,\ldots,L$. That is, $x_i = \sum_{j=1}^{L} x_i^j 2^{j-1}$, and $2^{L-1} \leq \max\{x_i\} < 2^L$. They want to determine whether all of their private secrets are equal with the assistance of the semi-honest TP. Note that there are two kinds of definition of the semi-honest TP. The first one, introduced by Chen *et al.* [16] for the first time, means that TP executes the protocol loyally and keeps a record of all its intermediate computations and might try to steal the parties' private secrets from the record, but he will not conspire with anyone. This means that the only way for this kind of TP to steal the parties' private secrets is to utilize the record of all the intermediate computations. And the second one, proposed by Yang *et al.*

[26], is that TP may misbehave on his own but will not conspire with anyone. This means that TP is allowed to try any possible means, including active eavesdropping strategy, to steal the parties' private secrets except conspiring with anyone. Up to now, the second kind definition of the semi-honest TP has been widely accepted as the most reasonable assumption. Therefore, our protocol adopts the second kind definition of the semi-honest TP.

All participants execute the following steps to complete the private comparison task.

***Preliminary:*** $P_i$ ( $i = 1, 2, 3$ ) generates $L$ two-level Bell state $|\Psi(0,0)\rangle = \frac{1}{\sqrt{2}} \sum_{j=0}^{1} |j\rangle|j\rangle$.

**Step 1:** TP prepares $L$ two-level four-particle cat states $|\Psi(u_0, u_1, u_2, u_3)\rangle = \frac{1}{\sqrt{2}} \sum_{j=0}^{1} (-1)^{ju_0} |j, j+u_1, j+u_2, j+u_3\rangle$, and arranges them into an ordered sequence

$$|\Psi(u_0^1, u_1^1, u_2^1, u_3^1)\rangle, |\Psi(u_0^2, u_1^2, u_2^2, u_3^2)\rangle, \ldots, |\Psi(u_0^L, u_1^L, u_2^L, u_3^L)\rangle, \tag{9}$$

where the superscripts denote the order of two-level four-particle cat states in the sequence. Then, TP takes particles with labels $(u_1^j, u_2^j, u_3^j)$ ( $j = 1, 2, \ldots, L$ ) out form each state $|\Psi(u_0^j, u_1^j, u_2^j, u_3^j)\rangle$ to construct new sequences labeled as

$$(u_1^1, u_1^2, \ldots, u_1^L), (u_2^1, u_2^2, \ldots, u_2^L), (u_3^1, u_3^2, \ldots, u_3^L), \tag{10}$$

which are denoted as $S_1, S_2, S_3$, respectively. Afterward, TP announces the ordered labels $(u_i^1, u_i^2, \ldots, u_i^L)$ to $P_i$ ( $i = 1, 2, 3$ ). Finally, for preventing eavesdropping, TP prepares three groups of decoy photons and inserts the $i$ th group into $S_i$ at random positions. Here, each decoy photon is one of the states randomly chosen from the set $V_1 = \{|k\rangle\}_{k=0}^{1}$ or $V_2 = \left\{ \frac{1}{\sqrt{2}} \left[ |0\rangle + (-1)^k |1\rangle \right] \right\}_{k=0}^{1}$. The new sequence of $S_i$ is denoted as $S_i'$. Finally, TP sends $S_i'$ to $P_i$.

**Step 2:** After confirming that $P_i$ ( $i = 1, 2, 3$ ) has received all particles, TP and $P_i$ check the security of the transmission of $S_i'$. Concretely, TP announces the positions and the bases of the decoy photons in $S_i'$ to $P_i$. According to the announced information, $P_i$ uses the bases as TP announced to measure the corresponding decoy photons and returns the measurement results to TP. Afterward, TP verifies these measurement results and checks whether eavesdroppers exist in the quantum channel. If there is no error, TP confirms that the quantum channel is secure and proceeds to the next step. Otherwise, they will abort this communication and restart the protocol.

**Step 3:** $P_i$ ( $i = 1, 2, 3$ ) discards the decoy photons in $S_i'$. Then, $P_i$ encodes her secret $x_i$.

For $i = 1, 2, 3$:
For $j = 1, 2, \ldots, L$:

$P_i$ sets $v_i^{'j} = x_i^j$ first. Then, $P_i$ generates the two-level Bell state $|\Psi(v_i^j, v_i^{'j})\rangle = \frac{1}{\sqrt{2}} \sum_{j=0}^{1} (-1)^{jv_i^j} |j\rangle |j + v_i^{'j}\rangle$ ( $v_i^j \in \{0,1\}$ ) by acting on $|\Psi(0,0)\rangle$ with $I \otimes |U_{(v_i^j, v_i^{'j})}\rangle$, i.e.,

$$|\Psi(v_i^j, v_i^{'j})\rangle = \left(I \otimes |U_{(v_i^j, v_i^{'j})}\rangle\right) |\Psi(0,0)\rangle, \tag{11}$$

where

$$|U_{(v_i^j, v_i^{'j})}\rangle = \sum_{j=0}^{1} (-1)^{jv_i^j} |j + v_i^{'j}\rangle\langle j|. \tag{12}$$

Afterward, $P_i$ performs the two-level Bell state measurement on the particle with label $u_i^j$ from the cat state and the particle with label $v_i^j$ from her Bell state, and knows the final state of particles with labels $(v_i^j - k_i^j, u_i^j - l_i^j)$. Consequently, $P_i$ can independently determine $k_i^j$ and $l_i^j$ with the knowledge of the labels of the Bell state she generated and the cat state's label $u_i^j$ announced by TP to her in step 1.

**Step 4:** For $j = 1, 2, \ldots, L$:
Three parties cooperate together to compute

$$S_L^j = \sum_{i=1}^{3} l_i^j, \quad S_K^j = \sum_{i=1}^{3} k_i^j. \tag{13}$$

Then, they announce $S_L^j$ and $S_K^j$ to TP.

**Step 5:** For $j = 1, 2, \ldots, L$:
At this stage, the $j$ th cat state is sent back to TP. Note that same to step 1, the decoy photons randomly chosen from the set $V_1$ or $V_2$ are used to guarantee the security of quantum transmissions. TP measures his state and obtains the labels

$$u_0^j + k_1^j + k_2^j + k_3^j, v_1^{'j} + l_1^j, v_2^{'j} + l_2^j, v_3^{'j} + l_3^j. \tag{14}$$

Then TP computes

$$S_C^j = u_0^j + \sum_{i=1}^{3}\left(k_i^j + v_i^{'j} + l_i^j\right). \tag{15}$$

By deducting $u_0^j + S_L^j + S_K^j$ from $S_C^j$, TP can get

$$S_V^j = \sum_{i=1}^{3} v_i^{'j}. \tag{16}$$

**Step 6:** If $S_V^j \bmod 3 = 0$ for all $j$, TP concludes that all the secrets of three parties are the same; otherwise, TP concludes that not all the secrets of three parties are the same. Finally, TP publicly tells $P_1, P_2, P_3$ the comparison result.

For clarity, the graphical description of entanglement swapping process of our protocol for three parties and for qubits is further given in Fig.2. Apparently, for $j = 1, 2, \ldots, L$, the two-level Bell state $|\Psi(v_i^j, v_i^{'j})\rangle$ generated by $P_i$ ($i = 1, 2, 3$) swaps entanglement with the original two-level four-particle cat state $|\Psi(u_0^j, u_1^j, u_2^j, u_3^j)\rangle$ according to formula (7).

### 3.2 The proposed MQPC protocol for $n$ parties and for qudits

In this subsection, we generalize the MQPC protocol for three parties and for qubits in Sect.3.1 to the version for $n$ parties and for qudits.

Suppose that there are $n$ parties named $P_1, P_2, \ldots, P_n$, where $P_i$ ($i = 1, 2, \ldots, n$) has a private secret $x_i$. The binary representation of $x_i$ in $F_{2^L}$ is $(x_i^1, x_i^2, \ldots, x_i^L)$, where $x_i^j \in \{0,1\}$ for $j = 1, 2, \ldots, L$. That is, $x_i = \sum_{j=1}^{L} x_i^j 2^{j-1}$, and $2^{L-1} \leq \max\{x_i\} < 2^L$. They want to determine whether all of their private secrets are equal with the assistance of the semi-honest TP. The second kind definition of the semi-honest TP is also adopted here.

All participants execute the following steps to complete the private comparison task.

***Preliminary:*** $P_i$ ($i = 1, 2, \ldots, n$) generates $L$ $d$-level Bell state $|\Psi(0,0)\rangle$.

**Step 1:** TP prepares $L$ $d$-level $n+1$-particle cat states $|\Psi(u_0, u_1, \ldots, u_n)\rangle$, and arranges them into an ordered sequence

$$\left|\Psi(u_0^1, u_1^1, \ldots, u_n^1)\right\rangle, \left|\Psi(u_0^2, u_1^2, \ldots, u_n^2)\right\rangle, \ldots, \left|\Psi(u_0^L, u_1^L, \ldots, u_n^L)\right\rangle, \tag{17}$$

where the superscripts denote the order of $d$-level $n+1$-particle cat states in the sequence. Then, TP takes particles with labels $(u_1^j, u_2^j, \ldots, u_n^j)$ ($j = 1, 2, \ldots, L$) out form each state $|\Psi(u_0^j, u_1^j, \ldots, u_n^j)\rangle$ to construct new sequences labeled as

$$\left(u_1^1, u_1^2, \ldots, u_1^L\right), \left(u_2^1, u_2^2, \ldots, u_2^L\right), \ldots, \left(u_n^1, u_n^2, \ldots, u_n^L\right), \tag{18}$$

which are denoted as $S_1, S_2, \ldots, S_n$, respectively. Afterward, TP announces the ordered labels $(u_i^1, u_i^2, \ldots, u_i^L)$ to $P_i$ ($i = 1, 2, \ldots, n$). Finally, for preventing eavesdropping, TP prepares $n$ groups of decoy photons and inserts the $i$ th group into $S_i$ at random positions. Here, each decoy photon is one of the states randomly chosen from the set $V_1$ or $V_2$. The new sequence of $S_i$ is denoted as $S_i^{'}$. Finally, TP sends $S_i^{'}$ to $P_i$.

**Step 2:** After confirming that $P_i$ ($i = 1, 2, \ldots, n$) has received all particles, TP and $P_i$ check the security of the transmission of $S_i^{'}$. Concretely, TP announces the positions and the bases of the decoy photons in $S_i^{'}$ to $P_i$. According to the announced information, $P_i$ uses the bases as TP announced to measure the corresponding decoy photons and returns the measurement results to TP. Afterward, TP verifies these measurement results and checks whether eavesdroppers exist in the quantum channel. If there is no error, TP confirms that the quantum channel is secure and proceeds to the next step. Otherwise, they will abort this communication and restart the protocol.

**Step 3:** $P_i$ ($i = 1, 2, \ldots, n$) discards the decoy photons in $S_i^{'}$. Then, $P_i$ encodes her secret $x_i$.

For $i = 1, 2, \ldots, n$:

For $j = 1, 2, \ldots, L$:

$P_i$ sets $v_i^j = x_i^j$ first. Then, $P_i$ generates the $d$-level Bell state $|\Psi(v_i^j, v_i^{'j})\rangle$ ($v_i^j \in \{0, 1, \ldots, d-1\}$) by acting on $|\Psi(0,0)\rangle$ with $I \otimes \left|U_{(v_i^j, v_i^{'j})}\right\rangle$, i.e.,

$$\left|\Psi\left(v_i^j, v_i^{'j}\right)\right\rangle = \left(I \otimes \left|U_{\left(v_i^j, v_i^{'j}\right)}\right\rangle\right) \left|\Psi(0,0)\right\rangle, \qquad (19)$$

where

$$\left|U_{\left(v_i^j, v_i^{'j}\right)}\right\rangle = \sum_{j=0}^{d-1} \zeta^{jv_i^j} \left|j + v_i^{'j}\right\rangle \langle j|. \qquad (20)$$

Afterward, $P_i$ performs the $d$-level Bell state measurement on the particle with label $u_i^j$ from the cat state and the particle with label $v_i^j$ from her Bell state, and knows the final state of particles with labels $\left(v_i^j - k_i^j, u_i^j - l_i^j\right)$. Consequently, $P_i$ can independently determine $k_i^j$ and $l_i^j$ with the knowledge of the labels of the Bell state she generated and the cat state's label $u_i^j$ announced by TP to her in step 1.

**Step 4:** For $j = 1, 2, \ldots, L$:
All parties cooperate together to compute

$$S_L^j = \sum_{i=1}^n l_i^j, \quad S_K^j = \sum_{i=1}^n k_i^j. \qquad (21)$$

Then, they announce $S_L^j$ and $S_K^j$ to TP.

**Step 5:** For $j = 1, 2, \ldots, L$:
At this stage, the $j$th cat state is sent back to TP. Note that same to step 1, the decoy photons randomly chosen from the set $V_1$ or $V_2$ are used to guarantee the security of quantum transmissions. TP measures his state and obtains the labels

$$u_0^j + k_1^j + k_2^j + \ldots + k_n^j, v_1^{'j} + l_1^j, v_2^{'j} + l_2^j, \ldots, v_n^{'j} + l_n^j. \qquad (22)$$

Then TP computes

$$S_C^j = u_0^j + \sum_{i=1}^n \left(k_i^j + v_i^{'j} + l_i^j\right). \qquad (23)$$

By deducting $u_0^j + S_L^j + S_K^j$ from $S_C^j$, TP can get

$$S_V^j = \sum_{i=1}^n v_i^{'j}. \qquad (24)$$

**Step 6:** If $S_V^j \bmod n = 0$ for all $j$, TP concludes that all the secrets of $n$ parties are the same; otherwise, TP concludes that not all the secrets of $n$ parties are the same. Finally, TP publicly tells $P_1, P_2, \ldots, P_n$ the comparison result.

The graphical description of entanglement swapping process of our protocol for $n$ parties and for qudits is further given in Fig.3. Apparently, for $j = 1, 2, \ldots, L$, the $d$-level Bell state $\left|\Psi\left(v_i^j, v_i^{'j}\right)\right\rangle$ generated by $P_i$ ($i = 1, 2, \ldots, n$) swaps entanglement with the original $d$-level $n+1$-particle cat state $\left|\Psi\left(u_0^j, u_1^j, \ldots, u_n^j\right)\right\rangle$ according to formula (7).

## 4  Analysis

As the MQPC protocol in Sect.3.1 is the special case of $n = 3$ for the MQPC protocol in Sect.3.2, we merely analyze the latter in detail here.

### 4.1  Output correctness

In this subsection, we show that the output of our protocol is correct. There are $n$ parties named $P_1, P_2, \ldots, P_n$, where $P_i$ ($i = 1, 2, \ldots, n$) has a secret $x_i$. The binary representation of $x_i$ in $F_{2^L}$ is $\left(x_i^1, x_i^2, \ldots, x_i^L\right)$, where $x_i^j \in \{0, 1\}$ for $j = 1, 2, \ldots, L$.

We take the $j$th bit of $x_i$ (i.e., $x_i^j$) for example, to illustrate the output correctness. $P_i$ encodes $v_i^{'j}$ (i.e., $x_i^j$) by producing $\left|\Psi\left(v_i^j, v_i^{'j}\right)\right\rangle$ through performing $I \otimes \left|U_{\left(v_i^j, v_i^{'j}\right)}\right\rangle$ on $\left|\Psi(0,0)\right\rangle$. Then, $P_i$ performs the $d$-level Bell state measurement on the particle with label $u_i^j$ from the cat state and the particle with label $v_i^j$ from her Bell state. As a result, the particle with label $u_i^j$ from the cat state and the particle with label $v_i^j$ from her Bell state swap entanglement. After $P_1, P_2, \ldots, P_n$ finish performing the $d$-level Bell state measurements, the $j$th cat state sent back to TP has the labels $u_0^j + k_1^j + k_2^j + \ldots + k_n^j, v_1^{'j} + l_1^j, v_2^{'j} + l_2^j, \ldots, v_n^{'j} + l_n^j$. TP computes $S_C^j$ and deducts $u_0^j + S_L^j + S_K^j$ from $S_C^j$ to obtain $S_V^j$. Apparently, we have

$$S_V^j = S_C^j - \left(u_0^j + S_L^j + S_K^j\right) = u_0^j + \sum_{i=1}^n \left(k_i^j + v_i^{'j} + l_i^j\right) - \left(u_0^j + S_L^j + S_K^j\right) = \sum_{i=1}^n v_i^{'j}. \qquad (25)$$

If $v_1^{'j} = v_2^{'j} = \ldots = v_n^{'j} = 0$, then $S_V^j = 0$; if $v_1^{'j} = v_2^{'j} = \ldots = v_n^{'j} = 1$, then $S_V^j = n$. These two cases imply that $S_V^j \bmod n = 0$. Otherwise, $S_V^j \bmod n \neq 0$. It can be concluded that the output of our protocol is correct.

### 4.2 Security

In this subsection, firstly, we show that the outside attack is invalid to our protocol. Secondly, we show that one party cannot learn other parties' secrets except for the case that their secrets are identical. The semi-honest TP who announces the comparison result cannot learn these parties' secrets either.

**(i) Outside attack**

We analyze the possibility for an outside eavesdropper to steal these parties' secrets according to every step of our protocol.

In our protocol, there are the qudit transmissions through the quantum channels in steps 1 and 5. An outside eavesdropper may utilize these qudit transmissions to extract some useful information about these parties' secrets by launching some famous attacks such as the intercept-resend attack, the measure-resend attack and the entangle-measure attack, *etc*. However, our protocol uses the decoy photon technique [46,47], which can be regarded as a variant of the effective eavesdropping check method of the BB84 protocol [1], to guarantee the security of the qudit transmissions. The effectiveness of decoy photon technology in 2-level quantum system towards the intercept-resend attack, the measure-resend attack and the entangle-measure attack has also been explicitly demonstrated in Refs.[48,49]. It is straightforward that the decoy photon technology adopted by our protocol is also effective towards these famous attacks in $d$-level quantum system. Therefore, an outside eavesdropper cannot steal any secret without being detected in steps 1 and 5.

In step 3, there is nothing transmitted. Hence, an outside eavesdropper can get nothing useful in this step.

In step 4, $P_1, P_2, \ldots, P_n$ announce $S_L^j$ and $S_K^j$ to TP. Even though an outside eavesdropper hears of $S_L^j$ and $S_K^j$, it is still helpless for her to steal any secret.

In step 6, TP publicly tells $P_1, P_2, \ldots, P_n$ the comparison result. An outside eavesdropper cannot know any secret in this step either.

It should be emphasized that the qudit transmissions are in a round trip in our protocol. As a result, the Trojan horse attacks from an outside eavesdropper including the invisible photon eavesdropping attack [50] and the delay-photon Trojan horse attack [51,52] should be taken into account. The way to prevent the invisible photon eavesdropping attack is that the receiver inserts a filter in front of her devices to filter out the photon signal with an illegitimate wavelength [52,53]. The way to prevent the delay-photon Trojan horse attack is that the receiver uses a photon number splitter (PNS:50/50) to split each sample quantum signal into two pieces and measures the signals after the PNS with proper measuring bases [52,53]. If the multiphoton rate is unreasonably high, this attack will be detected.

**(ii) Participant attack**

In 2007, Gao *et al.* [54] first pointed out that a dishonest participant's attack named participant attack is generally more powerful and should be paid more attention to. Up to now, the participant attack has attracted much attention in the cryptanalysis of quantum cryptography [55-57]. To see this in a sufficient way, we consider two cases of participant attack. Firstly, we discuss the participant attack from one or more dishonest parties, and then we analyze the participant attack from TP.

**Case 1: The participant attack from one or more dishonest parties**

Two situations should be considered. One is that one dishonest party wants to steal other parties' secrets; the other is that more than one dishonest parties conclude together to steal other parties' secrets. Note that TP is not allowed to conspire with any party.

**(a) The participant attack from one dishonest party**

Since the roles of $n$ parties are the same, without loss of generality, in this situation, we just consider the case that the dishonest $P_2$ wants to obtain the secret of $P_1$.

In our protocol, there is not any qudit transmitted between $P_1$ and $P_2$. If $P_2$ tries to intercept the transmitted particles from TP to $P_1$ in step 1 or from $P_1$ to TP in step 5, she will inevitably be caught as an outside eavesdropper in these two steps as analyzed above, since she has no knowledge about the positions and the bases of the decoy photons.

In step 3, $P_2$ can independently determine $k_2^j$ and $l_2^j$. In step 4, $P_2$ knows $S_L^j$ and $S_K^j$. However, her knowledge of $k_2^j, l_2^j, S_L^j$ and $S_K^j$ is still helpless for her to obtain $v_1^{'j}$, because $P_2$ has no chance to know $v_1^{'j} + l_1^j$.

It can be concluded that the dishonest $P_2$ cannot obtain the secret of $P_1$.

**(b) The participant attack from more than one dishonest parties**

Here, we only consider the extreme case of this situation that there are $n-1$ parties concluding together to steal the left party's secret, as in this extreme case the dishonest parties have the most power. Without loss of generality, assume that the dishonest $P_1, P_2, \ldots, P_{i-1}, P_{i+1}, \ldots, P_n$ collude together to obtain the secret of $P_i$.

Firstly, if any one of $P_1, P_2,...,P_{i-1}, P_{i+1},...,P_n$ tries to intercept the transmitted particles from TP to $P_i$ in step 1 or from $P_i$ to TP in step 5, she will inevitably be caught as an outside eavesdropper in these two steps as analyzed above, since she has no knowledge about the positions and the bases of the decoy photons.

Secondly, in step 3, $P_m$ $(m=1,2,...,i-1,i+1,...,n)$ can independently determine $k_m^j$ and $l_m^j$. In step 4, when they cooperate together, $P_1, P_2,...,P_{i-1}, P_{i+1},...,P_n$ can learn $l_i^j$ and $k_i^j$ from $S_L^j$ and $S_K^j$, respectively. However, they still cannot know $v_i^{'j}$, because they have no chance to know $v_i^{'j} + l_i^j$.

It can be concluded that the dishonest $P_1, P_2,...,P_{i-1}, P_{i+1},...,P_n$ cannot obtain the secret of $P_i$.

**Case 2: The participant attack from semi-honest TP**

Since TP is assumed as a semi-honest third party in our protocol, he may try his best to obtain the secret of $P_i$ without conspiring with any one. In step 4, TP receives $S_L^j$ and $S_K^j$. In step 5, TP knows the labels $v_i^{'j} + l_i^j$. However, he still cannot learn $v_i^{'j}$ due to the lack of knowledge on $l_i^j$, even though he knows $S_L^j$, $S_K^j$ and $v_i^{'j} + l_i^j$.

## 5 Discussion and conclusion

We compare our protocol with the previous MQPC protocols of Refs.[39-41] without considering the eavesdropping check process, as it can be thought as a standard procedure independent from the working principle of one quantum cryptography protocol. We show the comparison results in Table 1. Here, the qubit efficiency [8,58,59] is defined as

$$\eta = \frac{c}{t}, \quad (26)$$

where $c$ and $t$ are the number of compared classical bits and the number of consumed particles, respectively. Note that there are two MQPC protocols in Ref.[41], which are denoted as Refs.[41]-A and [41]-B in Table 1. It should be further emphasized that opposed to the MQPC protocol of Ref.[39] which is essentially insecure under the second kind definition of the semi-honest TP, our protocol is secure under this circumstance. Moreover, different from the protocol of Ref.[41]-B which needs the use of QKD method, our protocol need not adopt it to guarantee the security.

To sum up, in this paper, we propose a novel MQPC protocol based on the entanglement swapping of $d$-level cat states and $d$-level Bell states, where TP is allowed to misbehave on his own but will not conspire with any party. In our protocol, $n$ parties employ unitary operations to encode their private secrets and can compare the equality of their private secrets within one time execution of the protocol. Our protocol can withstand both the outside attacks and the participant attacks. None of the QKD methods is adopted to generate keys for security. One party cannot obtain other parties' secrets except for the case that their secrets are identical. The semi-honest TP, who knows the end comparison result on whether all private secrets from $n$ parties are equal, cannot learn any information about these parties' secrets either.

In addition, the practical implementation of our protocol may still be difficult at present, as the preparation of $d$-level $n+1$-particle cat states and $d$-level Bell states, their entanglement swapping, the $d$-level Bell state measurement and the $d$-level $n+1$-particle cat state measurement may still be hard to realize in actual situation with present quantum technologies. Thus, we just describe a theoretical feasible MQPC protocol in this paper. The practical realization of our protocol still needs great developments of quantum technologies in the future.

**Acknowledgements**

The authors would like to thank the anonymous reviewers for their valuable comments that help enhancing the quality of this paper. Funding by the National Natural Science Foundation of China (Grant Nos.61402407 and 11375152) is gratefully acknowledged.

**Compliance with ethical standards**

Conflict of interest: The authors declare that they have no conflict of interest.

**Appendix:**

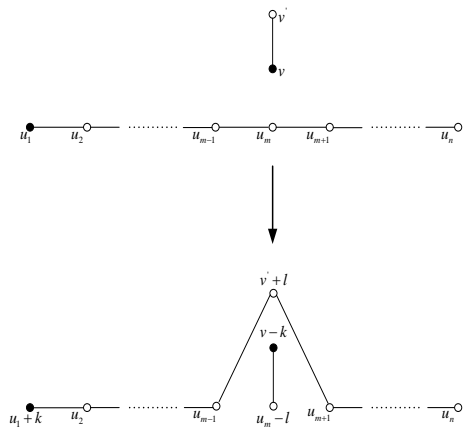

Fig.1 Graphical description of entanglement swapping between one $d$-level $n$-particle cat state and one $d$-level Bell state
(Here, one $d$-level $n$-particle cat state is depicted by a *line* with $n$ nodes on it where the first node is *solid* and other nodes are *empty*. And one $d$-level Bell state is depicted by a *line* with two nodes on it where one node is *solid* and the other node is *empty*.)

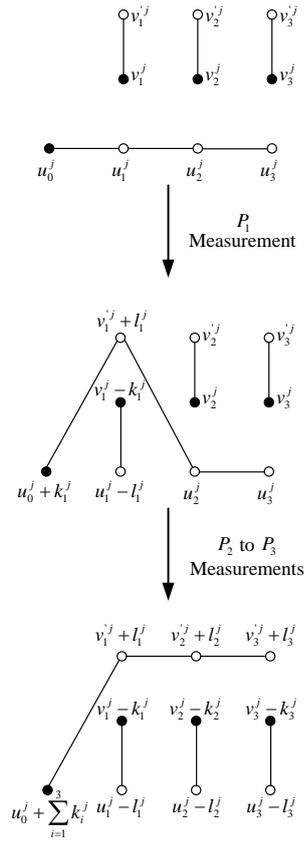

Fig.2 Graphical description of entanglement swapping process of our protocol for three parties and for qubits

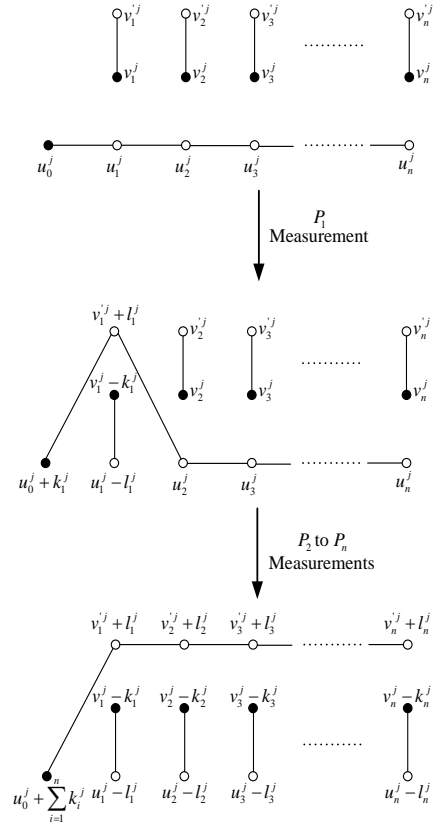

Fig.3 Graphical description of entanglement swapping process of our protocol for n parties and for qudits

Table 1 Comparison results between our protocol and the previous MQPC protocols

| | The protocol of Ref.[39] | The protocol of Ref.[40] | The protocol of Ref.[41]-A | The protocol of Ref.[41]-B | Our protocol |
|---|---|---|---|---|---|
| Quantum state | n-particle GHZ class state | d-level n-particle entangled state | d-level n-particle entangled state and d-level two-particle entangled state | d-level two-particle entangled state | d-level n+1-particle cat state and d-level two-particle Bell state |
| Quantum measurement for TP | No | d-level single-photon measurement | d-level single-photon measurement | d-level two-photon collective measurement | d-level n+1-particle cat state measurement |
| Quantum measurement for parties | Single-photon measurement | No | d-level single-photon measurement | No | d-level two-particle Bell state measurement |
| Unitary operation for TP | No | No | No | No | No |
| Unitary operation for parties | No | Yes | No | Yes | Yes |
| Quantum memory | No | No | Yes | Yes | Yes |
| Number of times of protocol execution | 1 (whether the jth group bits from any two parties are equal or not equal) | 1 (whether the jth group bits from n parties are all equal or not all equal) | 1 (whether the jth group bits from n parties are all equal or not all equal) | 1 (whether the jth group bits from n parties are all equal or not all equal) | 1 (whether the jth group bits from n parties are all equal or not all equal) |
| Quantum technology used | The entanglement correlation among different particles from one quantum entangled state | Quantum fourier transform | Quantum fourier transform | Quantum phase shifting operation | Quantum entanglement swapping |
| Qubit efficiency | $\frac{L}{nL}$ | $\frac{L}{nL}$ | $\frac{L}{3nL}$ | $\frac{L}{2L}$ (without considering the quantum resource used by a QKD protocol) | $\frac{L}{3nL+L}$ |